\documentclass[12pt,letterpaper]{article}
\usepackage[T1]{fontenc}
\usepackage{lmodern}
\usepackage[utf8]{inputenc}
\usepackage{geometry}
\usepackage{booktabs}
\usepackage{amsmath,amssymb,amsfonts,amsthm}
\usepackage{thmtools}
\usepackage{dsfont}
\usepackage{cancel}
\usepackage{bm}
\usepackage{mathtools}
\usepackage{changepage}
\usepackage{verbatim}
\usepackage{graphicx}
\usepackage{emptypage}
\usepackage{newlfont}
\usepackage[all,cmtip]{xy}
\usepackage[bottom]{footmisc}
\usepackage[titletoc,title]{appendix}
\usepackage{chngcntr}
\usepackage{apptools}
\usepackage{listings}
\usepackage{color}
\usepackage{sgame, tikz} 
\usepackage{pgfplots}
\usepackage{kpfonts}    

\AtAppendix{\counterwithin{lem}{section}}
\usepackage{caption}
\usepackage{subcaption}
\usepackage{float}
\usepackage{xr-hyper}
\PassOptionsToPackage{hyphens}{url}\usepackage[colorlinks=true,linkcolor=blue, allcolors=blue]{hyperref}
\usepackage{sgamevar}
\usepackage{lipsum}

\theoremstyle{plain} 
\newtheorem{theorem}{Theorem}
\newtheorem*{theorem*}{Theorem}

\newtheorem{lemma}{Lemma}
\newtheorem*{lemma*}{Lemma}

\newtheorem{proposition}{Proposition}

\theoremstyle{definition} 
\newtheorem{definition}{Definition}
\newtheorem*{definition*}{Definition}

\newtheorem*{prop*}{Properties}

\newtheorem*{ex*}{Example}

\newtheorem{claim}{Claim}

\newtheorem{assumption}{Assumption}

\newcommand{\argmax}{\operatornamewithlimits{argmax}}

\usepackage{setspace}
\reversemarginpar

\usepackage[nameinlink]{cleveref}
\crefname{manualasm}{assumption}{assumptions}
\crefname{innercustomthm}{theorem}{theorems}
\crefname{prop}{proposition}{propositions}
\crefname{ex}{example}{examples}
\crefname{defn}{definition}{definitions}
\crefname{claim}{claim}{claims}
\crefname{lem}{lemma}{lemmas}
\crefname{thm}{theorem}{theorems}

\newcommand{\sigmalr}{\sigma_{L \to R}}

\allowdisplaybreaks
\usepackage[shortlabels]{enumitem}

\usepackage{natbib}

\pgfplotsset{compat=1.17}

\title{Rediscovery}
\author{Martino Banchio\thanks{Università Bocconi, IGIER, and Google Research. Email: \href{mailto:martino.banchio@unibocconi.it}{martino.banchio@unibocconi.it}}
\and
Suraj Malladi\thanks{Northwestern Kellogg School of Management. Email: \href{mailto:suraj.malladi@kellogg.northwestern.edu}{suraj.malladi@kellogg.northwestern.edu}}
}

\date{}
\pgfplotsset{compat=1.15}
\begin{document}
\maketitle

\begin{abstract}
We model search in settings where decision makers know what can be found but not where to find it. A searcher faces a set of choices arranged by an observable attribute. Each period, she either selects a choice and pays a cost to learn about its quality, or she concludes search to take her best discovery to date. She knows that similar choices have similar qualities and uses this to guide her search. We identify robustly optimal search policies with a simple structure. Search is directional, recall is never invoked, there is a threshold stopping rule, and the policy at each history depends only on a simple index.
\end{abstract}

\section{Introduction}

Making an original discovery (e.g., tackling an open problem or developing a breakthrough innovation) is a process of trial and error in the face of stark uncertainty. Researchers, agencies and firms learn from their past successes and dead-ends when deciding which approach to try next. They also infer from their attempts whether worthwhile discoveries even exist and decide when to give up.

On the other hand, searchers attempting rediscovery know that worthwhile discoveries exist, even if they do not know where to find them. A student tackling a homework problem faces much of the same uncertainty as the researcher who first solved it. However, the student knows that the problem has a solution and relies only on material covered in class, whereas the researcher had no such guarantees. Analogously, a non-nuclear state faces significant uncertainty when trying to develop a weapon, but they know that other programs --- starting with the Manhattan project --- succeeded in the same endeavor.

Intuitively, the process of rediscovery seems simpler than original discovery. For example, while comparing OpenAI to its imitators, OpenAI's CEO Sam Altman argues: 
\begin{quote}
    It's really easy to copy something you know works. One of the reasons people don't talk about why it's so easy is that you have the conviction to know that it's possible. And so after a research lab does something, even if you don't know exactly how they did it, it's --- I won't say easy, but --- doable to go off and copy it [...] What is really hard [...] is the repeated ability to go off and do something  new and totally unproven... A lot of organizations talk about the ability to do this. There are very few that do, across any field.\footnote{This quote is from the \href{https://www.youtube.com/watch?v=peg-aX1oii4&t=963s}{Sam Altman} episode of 20VC with Harry Stebbings.} 
\end{quote}

We develop a stylized model of rediscovery and characterize the optimal search process.
Our model crystallizes why the process of rediscovery may feel ``easy'' in many respects.
While original discovery involves learning where to look and inferring what can be found, rediscovery only involves the former.
With original discovery, a bad outcome can be a sign to stop searching: perhaps all feasible approaches lead nowhere.
But with rediscovery, a bad outcome is always a sign to continue, albeit with a sufficiently different approach, because good outcomes exist.
It is also a sign for the searcher to aspire for more than before stopping, because there are fewer approaches left to try.

Understanding rediscovery may be of interest to economists, because it appears to be a ubiquitous and important driver of innovation. Firms that develop novel technologies, such as self-driving cars or new AI algorithms, often keep their methods as trade secrets instead of publicizing them or filing a patent. Competitors who learn only that such inventions are feasible may embark on their own R\&D process to at least partially recreate the innovating firm's success. In the terms of this example, we ask: how do these competitors go about searching the space of possible designs? At what point do they conclude search and release their own novel variants of the original innovation?

Similar questions have been considered in the management and entrepreneurship literature on the behavioral theory of the firm. In a seminal book, \citep{cyert1963} describe \emph{problemistic search} by managers which is triggered by events like ``failure to achieve the profit goal'' or ``innovation by a competitor.'' Subsequent authors have suggested that this sort of problemistic search by firms often happens over \textit{rugged landscapes}, meaning that the mapping from firm’s choices to outcomes is complex and unpredictable \citep*{levinthal1997adaptation, billinger2014search, callander2011searching, callander2022innovation}. Motivated by this literature, we conceive of rediscovery as problemistic search over rugged landscapes. Whereas much of the economics literature has focused on how social learning \textit{across} firms affects innovation over a rugged landscape, we adapt \citet{malladi2022searching} to capture the innovation process \textit{within} a rational, forward-looking firm.

\paragraph{Model and Results}
In our model, there is a continuum of choices between 0 and 1 that have unknown payoffs.  A searcher can learn the payoff to any given choice at a cost. Each period she decides whether to continue searching, and if so, which choice to learn about next. She eventually stops to take the best choice she had discovered so far.

Crucially, the searcher knows that there exists some choice which achieves a certain target payoff (e.g., that it is possible to design an invention of a given quality), but she does not know which. We interpret this as capturing rediscovery: the searcher knows with certainty, perhaps by seeing a predecessor's success, that a good discovery is possible, but she does not know \textit{a priori} where to find it. The searcher also knows that the mapping from options to payoffs is Lipschitz continuous with a known Lipschitz constant. This assumption captures the idea that the searcher explores a rugged landscape, as she entertains a rich set of possibilities about the shape of mapping from options to payoffs. The assumption simultaneously captures learning from past failures and successes through trial and error. The searcher's past discoveries guide where she looks next, as continuity implies that proximate choices yield similar payoffs.

The searcher's utility upon concluding search is the payoff of her best discovery minus the sum of her accumulated search costs.
We assume that the searcher follows a plan that, at every history, maximizes her \textit{worst-case} utility upon concluding search. That is, she searches in a way that is robust to the shape of the complex and unpredictable rugged landscape. At each history, and for every plan of search, she evaluates this worst case over the possible shapes of the mapping from choices to payoffs, knowing only that this mapping must be Lipschitz continuous, pass through the points she had previously discovered, and attain the benchmark payoff somewhere.

We find an optimal search policy that is simple in many ways. 

First, the searcher follows a threshold stopping rule, meaning she stops if the payoff she discovers exceeds that period's threshold and continues otherwise. Moreover, these thresholds increase with time: the knowledge that good discoveries exist somewhere causes the searcher to become emboldened rather than discouraged by bad discoveries.

Next, search proceeds from left to right, although the searcher is free to look anywhere in the search space at any time. The knowledge that good discoveries exist makes search a process of elimination. Search can be performed methodically, ruling out unfruitful regions of the search space and honing in on the location of more promising choices.

Third, the searcher has perfect recall but never invokes it. She always selects the last option she had discovered rather than returning to a previous discovery she had made. 

Finally, while search happens in a non-stationary environment, the optimal search policy depends only on a simple index. We define the \textit{search window} at a given history to be the set of choices which can potentially achieve the benchmark quality that the searcher would ideally rediscover. Both the optimal search and stopping rule are pinned down by the length of the search window.

None of these properties hold for the optimal policies identified in \citet{malladi2022searching}, which studied search in a similar setting but when good discoveries are not guaranteed to exist. The comparison illustrates how rediscovery is procedurally simpler than search for original discovery.

\paragraph{Related Literature}
Our model contributes to the literature on search theory. Early papers treat search as a pure stopping problem \citep{mccall1970economics, rothschild1974searching}. \citet{weitzman1979optimal} considers \textit{ordered search} over a set of independent but not identically distributed items, where searchers can select which item to explore at each history and when to stop. 

Relaxing independence to allow cross-item learning is notoriously difficult, and a solution is known only in a special case of conditionally independent items \citep{adam2001learning}. \citet{callander2011searching} and \citet{garfagnini2016social} capture richer learning by modeling the mapping from items to payoffs as the realized path of a Brownian motion. They, and a subsequent literature, study ordered search by a sequence of short-lived searchers.\footnote{See \cite{callander2019risk, callander2022innovation, carnehl2025quest}.} 
By contrast, our model follows \citet{malladi2022searching} in solving for forward-looking ordered search by a long-lived searcher.\footnote{In particular, we study sequentially robust search policies. For other recent perspectives on robustness and dynamics, see \cite{libgober2022coasian} and \cite{auster2024prolonged}.} 

The Brownian framework has also been used to capture learning in settings beyond sequential ordered search, which is the focus of our paper. Notably, \cite{Urgun2024} and \cite{Wong2024} study optimal contiguous search and experimentation over a Brownian path, where searchers choose how quickly to explore. By contrast, we study a searcher who can freely choose where to explore each period. \cite{bardhi2024attributes} and
\cite{bardhi2023local} study optimal information acquisition (by a single agent or by delegation to several agents, respectively) about a complex project with correlated attributes. Where these models are static, we study a dynamic model in which optimal sequential search is markedly different from simultaneous search.

Our paper is also related to the topics of bandits and optimization in computer science and operations research \citep{lattimore2020bandit, hansen1992global}. A key difference is that we study rational---forward-looking and dynamically consistent---searchers. Moreover, our searchers face a stopping problem, which affects optimal exploration.

\section{Model}\label{sec:model}

There is a continuum of items arranged along the interval $S \equiv [0, 1] \subset \mathbb{R}$.
Let $Q \subset [0, 1]\to \mathbb{R}$ be the set of potential \textit{quality indices}---mappings from the search space to a measure of quality. There is some true quality index $q \in Q$, so each item $x \in S$ has a quality $q(x) \in [0, 1]$.

There is an agent who knows $Q$ but not the true quality index.
She can learn the quality of items in $[0, 1]$ through costly search. This way, she narrows down the set of candidate true quality indices in $Q$.

In each period, $t=0, 1, 2, 3 \ldots$, the agent takes one of two kinds of actions. 
She either explores a new item $x_t \in [0, 1]$ to learn its quality, $q(x_t)$. 
Or she concludes her search, $x_t = \emptyset$,
and adopts the highest quality item that she had discovered so far, including an outside option of quality zero.

Formally, let $h_t = \{(x_i, z_i)\}_{i=0}^{t-1}$ be the time $t$ partial history when the agent has not yet concluded search, with $z_i = q(x_i)$. 
Let $H$ denote the set of all partial histories.
Let $X_{h_t}$ be the set of items that were explored at $h_t$. 
Let $z^*_{h_0} = 0$ be the outside option, and for $t\geq 1$, $z^*_{h_t} = \max \{0,z_0,\dots,z_{t-1}\}$. If $x_i \in X_{h_t}$ is such that $z_i = z^*_{h_t}$, then $x_i$ is a \textit{best item at $h_t$}. 

A quality index $\tilde{q} \in Q$ is \textit{consistent} at $h_t$ if $\tilde{q}(x_i) = z_i$ for all $i=0, \ldots, t-1$.
Let $Q_{h_t} \subset Q$ be the set of consistent quality indices at $h_t$. 

We assume the following throughout:

\begin{assumption}\label{assumption}
    $Q$ is the set of all $L$-Lipschitz continuous mappings $q: [0, 1] \to \mathbb{R}$ such that $q(x) = 1$ for some $x \in [0, 1]$.
\end{assumption}

In essence, the agent knows little about the shape of the true quality index. She knows that proximate items in $[0, 1]$ cannot be too different in quality. She also knows that there exists some item of at least a certain quality, so search may be worthwhile.  But she does not know a priori where to find such an item. We call this known achievable quality the \emph{quality standard} and normalize it to 1. Items that achieve the quality standard are \emph{targets}.

\subsection{Payoffs}\label{sec:payoffs}
The agent’s benefit to adopting item $x$ is $q(x)$. The agent’s cost of exploring item $x$ in any period is $c>0$. The agent's total payoff at history $h_{t+1} \in \tilde{H}$  such that $x_t = \emptyset$ is given by:
$$p(h_{t+1}) = z^*_{h_{t+1}} - c \cdot t.$$
That is, when an agent stops at history $h_t$, she will adopt an item of quality $z^*_{h_{t+1}}$. 

\subsection{Strategies and Policies}

A strategy of the agent is a deterministic mapping $\sigma \colon H \to \big\{ [0, 1]\cup \{\emptyset\} \big\}$. A strategy $\sigma$ \textit{eventually terminates} if for all $h \in H$ and $\tilde{q}\in Q_h$, $\sigma$ reaches a terminal history from $h$ when $ q = \tilde{q}$. We restrict attention to the set of all strategies $\Sigma$ that eventually terminate.

We denote by $h^{+1}_q(\sigma)$ the history that follows $h$ if the agent adopts policy $\sigma$ and the quality index is $q$, i.e. 
\[h^{+1}_q(\sigma) = h \cup \Big\{\big(\sigma(h), q(\sigma(h))\big) \Big\},\]
and similarly $h^{+2}_q(\sigma)$, $h^{+3}_q(\sigma)$, etc. The set of reachable histories for a given quality index $q$ is then $H^\sigma_q = \{h_0, (h_0)^{+1}_q, (h_0)^{+2}_q,\dots \}$.\footnote{We drop the dependence on $\sigma$ when the strategy is clear from the context.} We denote by $H^\sigma \subset H$ the set of histories that strategy $\sigma$ can reach along her decision tree (i.e., for some  $q \in Q$) starting from the empty history. Formally,  $H^\sigma = \{H^\sigma_q : q \in Q\}$.
A \textit{search policy} $\sigma|_D$ is a restriction of $\sigma$ to some domain $D \supset H^{\sigma}$. A search policy contains sufficient information to describe how search unfolds and when it stops for any $q \in Q$, because it at least specifies actions for reachable histories. A search policy \emph{terminates} if, for any $q$ there exists $h \in H^\sigma_q$ such that  $\sigma(h) = \emptyset$. Such a history $h$ is called \emph{terminal}.

\subsection{Objective}\label{sec:objectives}
To capture the unpredictability about the shape of the true quality index (i.e, the idea of rugged landscapes), we take the view that the agent does not have a prior over $Q$. 
She seeks a strategy that maximizes the eventual payoff that she is guaranteed, starting from \textit{any} history and regardless of which consistent quality index at that history is realized.

{
Under a strategy $\sigma \in \Sigma$ and starting from a history $h$, a quality index $q\in Q_h$ induces a terminal history $h^\sigma_q$ and its corresponding terminal payoff $p(h^\sigma_q)$. 
A strategy $\sigma^*$ is \textit{optimal} if at the empty history $h=h_0$
\[
\sigma^* \in \argmax_{\sigma \in \Sigma} \Big\{ \min_{q \in Q} p(h^\sigma_q)\Big\}.
\]
Similarly, a search policy $\sigma^*|_D$ is optimal if it can be extended to a strategy $\sigma^*$ that satisfies the above condition.
}

\subsection{Dynamic Consistency}
The agent can be thought of as choosing a fully contingent plan at time zero that maximizes her worst-case eventual payoff upon stopping. 
A natural question is: would the agent stick to her \textit{ex-ante} optimal plan at every later history even if she were given the chance then to revise it? 

\begin{definition}\label{definition:dynamic consistency}
    A policy $\sigma$ is \emph{dynamically consistent} if at all histories $h \in H^\sigma$
    \[
    \sigma \in \argmax_{\sigma \in \Sigma} \Big\{ \min_{q \in Q_h} p\big(h^\sigma_q\big)\Big\}
    \]
\end{definition}

A dynamically consistent strategy is also optimal, as Definition \ref{definition:dynamic consistency} holds, in particular, at the empty history.
In addition to being optimal, a dynamically consistent policy maximizes the agent's worst-case eventual payoff after any reachable history, where worst-case is taken over the consistent quality indices at that history. In words, an agent following a dynamically consistent policy would not revise it at any reachable history, even if given the chance. 

Optimality and dynamic consistency coincide in models that assume Bayesian updating and subjective expected utility, but the equivalence need not hold in preference models with ambiguity aversion.\footnote{See \cite{hanany2009updating} for a characterization of update rules for which dynamic consistency holds in models with ambiguity aversion.}
For example, \citet{auster2024prolonged} study a stopping problem in the context of information acquisition with ambiguity. They find an optimal policy that is dynamically inconsistent, exhibiting non-monotonicity in beliefs and randomized stopping.

On the other hand, \citet{epstein2003} show that prior-by-prior updating of the decision maker's beliefs and \emph{rectangularity} of the set of priors ensures dynamic consistency. These conditions hold in our model, taking the set of priors to be set of dirac measures over the space of consistent quality indices at some history and the set of posteriors as the dirac measures over consistent quality indices at a subsequent history.\footnote{Rectangularity requires that any combination of marginal beliefs with \emph{any} conditional probability distribution is a plausible prior. Marginal beliefs in our model are Dirac measures, so every such combination is itself a Dirac measure.}

To capture rational behavior where the agent is not in conflict with her future self, we look for dynamically consistent policies.

\section{An Optimal Search Policy}\label{sec:opt_search}
In this section we characterize an optimal search policy in closed form. Our main result shows that, even though the set of feasible policies for the agent is rich, and histories can be quite complex, there exist optimal policies that take a fairly simple form.

\subsection{Simple Policies}\label{sec:simple}

We begin by introducing certain classes of simple search policies. 

\begin{definition}
A policy $\sigma$ is a (left-to-right) \textit{directional policy} if for every $h \in H^\sigma$, either $\sigma(h) = \emptyset$ or $\sigma(h) > x$ for all $x\in X_h$.     
\end{definition}

In words, a directional policy is one where the agent searches along one direction in the search space rather than bouncing back and forth. Note that fixed search direction might reasonably capture a shopper walking though the aisles of a grocery store or a pharmaceutical company experimenting incrementally with drug dosages. Taking such examples as motivation, some models assume directionality as a natural constraint on the space of search strategies (e.g., \cite{arbatskaya2007ordered, Urgun2024, Wong2024}).

\begin{definition}
    A policy $\sigma$ is a \textit{threshold policy} if, for every non-terminal history $h \in H^\sigma$, there exists a $\tau_{h}$ such that 
    $\sigma(h^{+1}_q) = \emptyset$ if and only if $z^*_{h^{+1}_q} \geq \tau_h$.    
\end{definition}

A policy is a threshold policy if, prior to searching, the agent has a threshold in mind such that if search yields a quality exceeding that threshold she will conclude search, and she will otherwise continue searching. While threshold stopping rules are typically optimal in simple search models where agents take independent draws from a known distribution, they need not be optimal when there is learning. For example, an agent may stop after a sufficiently \textit{bad} draw if this discourages her about the prospect of making good discoveries (e.g., see \cite{rothschild1978searching, malladi2022searching, Urgun2024}). 

\begin{definition}
    A policy $\sigma$ \textit{ignores past discoveries} if $\sigma(h_t) = \emptyset$ implies $z^*_{h_t} = z_{t-1}$, i.e. the agent always takes the last item discovered
\end{definition}

In the model we assume the agent has perfect recall and always takes the highest quality item discovered to date. Therefore a strategy can only ignore past discoveries if the agent continues searching until her best discovery was her last.\footnote{Note that having the threshold property does not imply that the agent ignores past discoveries. She can, for example, stop regardless of the outcome of her second search (i.e., a threshold stopping rule with a threshold of zero) and take the first discovery if her second one is worse.} Recall can be valuable in contexts with learning, as bad draws can cast previous good discoveries in better light (e.g., see \cite{malladi2022searching, Urgun2024}).

Next, note that histories are complex and their dimensionality increases with time. Optimal policies may depend intricately on the sequence of past realizations. Here, we define a class of policies that, instead, depend on a simple one-dimensional state variable of any history. 

To that end, we introduce the notion of a \emph{search window} at history $h$:
\[S_{h} \equiv \big\{x \in [0,1]\ \big| \ \exists q \in Q_{h} \text{ s.t. } q(x) = 1\big\}.\]
The search window is the set of items $x$ which are targets under some consistent quality index $q \in Q_{h}$. Alternatively, if we denote by $\overline{q}_{h}: [0,1] \to \mathbb{R}$ the upper-envelope of feasible quality indices at history $h$,
then $S_{h}$ is the set of items $x$ for which $\overline{q}_{h}(x) \geq 1$.

\begin{figure}[!h]
\centering
\begin{subfigure}{0.3\textwidth}
    \centering 
    \includegraphics[width=46mm]{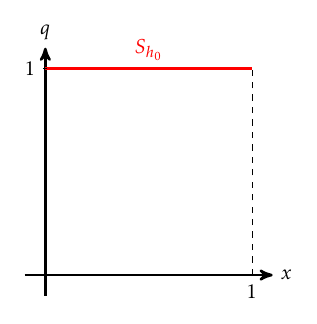}
    \caption{}
    \label{fig:searchwindow0}
\end{subfigure}
\hspace{.8em}
\begin{subfigure}{0.3\textwidth}
    \centering    
    \includegraphics[width=46mm]{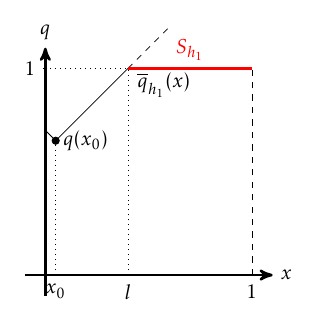}    
    \caption{}
    \label{fig:searchwindow1}   
\end{subfigure}
\hspace{.8em}
\begin{subfigure}{0.3\textwidth}
    \centering 
    \includegraphics[width=46mm]{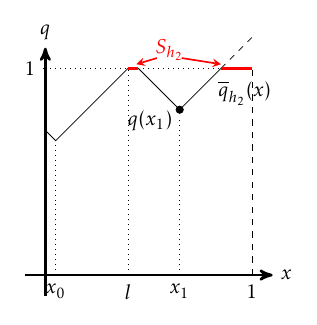}
    \caption{}
    \label{fig:searchwindow2}   
\end{subfigure}
 \caption{The red line represents the search window. The solid black line instead represents $\overline{q}_{h_t}$, the upper envelope of qualities given the current history. }
\label{fig:first choice}
\end{figure}

The search window shrinks with additional searches, so $S_{h^{+1}_q} \subseteq S_{h}$ for any $q\in Q_h$.\footnote{Observe that the set of feasible quality indices shrinks with additional searches, i.e.,  $Q_{h^{+1}_q} \subseteq Q_{h}$, and therefore $\overline{q}_{h} \geq \overline{q}_{h^{+1}_q}$ everywhere.} At any history $h$ where $(x, z) \in h$ and $z<1$, the open interval of length $\frac{2(1-z)}{L}$ centered at $x$ lies outside of the $S_{h}$.
More generally, at history $h = \{(x_0, z_0), \dots, (x_t, z_t)\}$,  $$S_{h} = [0, 1] - \bigcup_{j = 0}^t \big(x_j - \frac{(1-z_j)}{L}, x_j + \frac{(1-z_j)}{L}\big).$$ This is depicted in  \Cref{fig:first choice}.

Let $l_h \equiv ||S_h||$, i.e., the lebesgue measure of the search window, for all $h \in H$. 

\begin{definition}
    A policy $\sigma$ is an \textit{index policy} if $\sigma$ is measurable with respect to the lebesgue measure of the search window on $H^\sigma$. 
\end{definition}

Index strategies are those which depend on histories only through the length of the search window. This is particularly attractive as the unique histories grow instead exponentially in the number of searches. Optimal search behavior can be fully characterized in terms of a single-dimensional ``sufficient statistic'' of the exponentially-large state space.

\subsection{Main Result}\label{sec:main_result}
Our main result is that:

\begin{theorem}
There exists an optimal policy which is directional, threshold, index and ignores the past. Furthermore, this policy is dynamically consistent.
\end{theorem}

\noindent
The remainder of this section is dedicated to constructing such a policy explicitly.
We fix a Lipschitz constant $L=1$ for ease of notation.  

An \emph{ordered search history} is a history $h$ where the search window is an interval of the form $[a,1]$ for some $a\in[0, 1]$. \Cref{fig:first choice} shows search windows at three histories: the first two are ordered search histories, because $S_{h_0}$ and $S_{h_1}$ are intervals that contain the endpoint 1. The third history is not an ordered search history, as $S_{h_2}$ is disconnected.

We define a search policy $\sigma_{L\to R}$ in which the agent explores $S$ from left to right and stops whenever she makes a discovery exceeding an increasing, history-dependent threshold. To this end, we introduce two auxilliary functions.

Let $N\colon [0,1]^2 \to \mathbb{N}$ be defined as follows:
\begin{equation}\label{eq:N(c,a)}
 N(c,l) \equiv \begin{cases}
    0 \quad \text{ if } c \in \Big(1-\frac{l}{2},1\Big], \\
    1 \quad \text{ if } c \in  \Big(\frac{l}{2},  1-\frac{l}{2}\Big],\\
    n \quad \text{ if } c \in \Big(\frac{l}{n(n+1)},\frac{l}{n(n-1)}\Big].
\end{cases}   
\end{equation}
Roughly, $N$ maps search costs and the length of a search window to the maximum number of searches that $\sigma_{L\to R}$ makes for any $q \in Q$. In keeping with this interpretation, $N$ is decreasing in costs. Next, when costs of search are sufficiently low, $N$ is increasing with interval length: more space left to explore means more searches might be needed to discover a good quality item. But when costs are sufficiently high, $N$ is decreasing with interval length increases: more space left to explore discourages the agent from exploring at all. 

When $N(c, l) \neq 0$, define $\phi \colon [0, 1] \to \mathbb{R}$ as 
\begin{equation}\label{eq:phi(l)}
    \phi(l) = 1-\frac{l}{2N(c,l)} - \frac{N(c,l)-1}{2}c
\end{equation}
The function $\phi$ maps the length of a search window to a quality threshold that is used to define the stopping region in $\sigmalr$. It is straightforward to check that $\phi$ is decreasing. As the search window grows larger, more search is potentially required to find a good outcome. Therefore, the agent is willing to conclude search for lower quality discoveries. When the remaining space to be seached is small, further search is likely to secure items close to the benchmark quality, so the threshold for stopping is higher.

\begin{definition}\label{definition:opt_policy} 
    The \emph{left-to-right} search policy $\sigmalr\colon H^{\sigmalr} \to \Delta\{S\cup \emptyset\}$ is given by
    \begin{align*}
        \sigmalr(h) &= \begin{cases}
        \emptyset \hspace{8.8em}  \text{ if } z^*_{h} \geq \phi(l_h) - c,  \\
        1-l_{h}+ 1 - \phi(l_{h}) \hspace{2.5em} \text{ otherwise,}
    \end{cases}
    \end{align*}
    for all $h \in H^{\sigmalr}$.
\end{definition}

\begin{figure}
    \centering
    \includegraphics[width=0.9\linewidth]{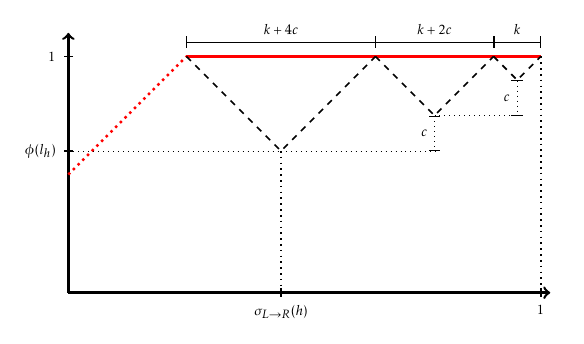}
    \caption{Visualizing $\sigmalr$ at some ordered search history $h$, where the search window is given by the solid red line.}
    \label{figure:triangles}
\end{figure}

\Cref{figure:triangles} shows that $\sigmalr$ has a simple geometric characterization at any ordered search history $h$. Let $k\leq 2c$ be the largest real number such that the search window $S_h$ can be partitioned in balls of diameter $k, k+2c, k+4c, \dots$.
The number of such balls is $N(c,l_h)$. Order these balls from the largest to the smallest in the search window. The left-to-right policy searches at the center of the largest ball at history $h$ and stops if and only if $q(\sigmalr(h))$ attains a value of at least the peak of the corresponding triangle. 

Although $\sigmalr$ is well-defined at all histories, the left-to-right policy stops at the first history that is not ordered.

\begin{lemma}\label{lemma:terminal or ordered}
For any $q\in Q$ and non-terminal $h \in H^{\sigmalr}$, exactly one of the following conditions is satisfied:
    \begin{enumerate}
        \item $h^{+1}_q(\sigmalr)$ is a terminal history, or
        \item $h^{+1}_q(\sigmalr)$ is an ordered search history.
    \end{enumerate}
\end{lemma}
The proof of this result, together with all other omitted proofs, is in \Cref{appendix:proofs}.
\Cref{fig:windows} gives intuition for \Cref{lemma:terminal or ordered}. Suppose search reveals that $q(x) > \phi(l)$, as in \Cref{fig:orderedsearch_split}. 
The history following this observation is no longer an ordered search history, but $\sigmalr$ stops here. Instead, suppose that the agent observes a quality $q(x)$ lower than the threshold, as in \Cref{fig:orderedsearch_window}. Then, the history that follows is an ordered search history, and $\sigmalr$ determines what to do next.
\begin{figure}[h]
    \centering
    \begin{subfigure}{0.47\textwidth}
    \hspace*{-.6cm}
    \includegraphics{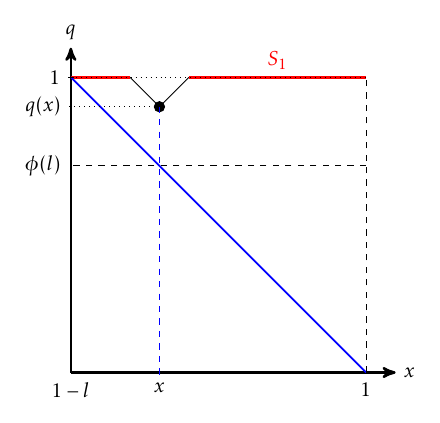}
    \caption{}
    \label{fig:orderedsearch_split}
    \end{subfigure}%
    \begin{subfigure}{0.47\textwidth}
    \hspace*{-.5cm}
    \includegraphics{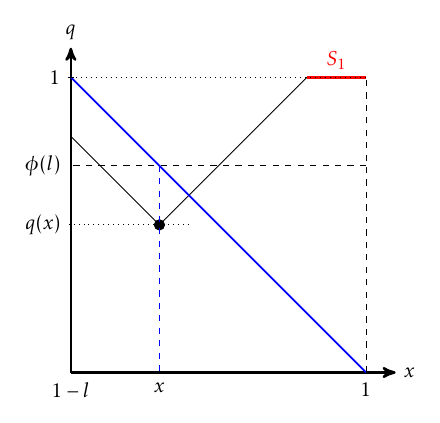}
    \caption{}
    \label{fig:orderedsearch_window}
    \end{subfigure}
    \caption{}
    \label{fig:windows}
\end{figure}

\begin{proposition}\label{prop:sigmalr is simple}
The left-to-right policy $\sigmalr$ is directional, threshold, index, and ignores past discoveries.
\end{proposition}

From the definition of $\sigmalr$, it immediately follows that this is a \textit{threshold} and \textit{index} policy. 

By \Cref{lemma:terminal or ordered}, a non-terminal, reachable history is an ordered search history as well. As shown in \Cref{fig:orderedsearch_window}, such a history has a smaller search window than the one before it. Because $\sigmalr$ always searches inside the shrinking search window of ordered search histories on path, it is \textit{directional}. 

Because $\phi$ is decreasing in the length of the search window, the stopping threshold increases on path. Therefore, the agent \textit{ignores past discoveries}, as she only stops if she exceeds an ever-higher target.

\begin{theorem}\label{thm:opt}
 $\sigmalr$ is an optimal, dynamically consistent search policy.
\end{theorem}

We discuss the intuition behind \Cref{thm:opt} in \Cref{sec:two periods} and \Cref{sec:general}, while a formal proof is in \Cref{appendix:proofs}.

\subsection{Discussion}

The optimal search policy bears some resemblance to the observations made in the literature on problemistic search and rugged landscapes.

First, we find that the agent is never discouraged by a poor payoff draw, because she knows good discoveries exist. By contrast, she becomes more ambitious and sets a higher stopping threshold for her next search. Therefore, the agent always takes the last item she had discovered when she finally stops. The idea that agents attempting rediscovery are not easily discouraged corresponds with the observation by \citet{cyert1963} that, in problemistic search by firms and managers, \emph{``[so] long as the problem is not solved, search will continue.''}
This contrasts with the behavior characterized in the literature on reference points and aspiration formation (e.g. see \citet{Dalton2018}, \citet{selten1998}, \citet{karandikar1998}), where unrealized aspirations negatively affect utility and are revised downward. 
With rediscovery, the quality standard is known to be achievable, so aspirations are never revised.

On the other hand, the agent in our model does not begin to explore the space unless the quality standard lies sufficiently higher than the status quo. 
It reflects the finding by \citet{baum2007} that \emph{``[performance] below aspirations [...] leads decision makers to initiate experimentation to identify new ways of doing things and new things to do, while satisfactory performance does not.''}

\section{Analysis of the Two Period Case}\label{sec:two periods}

We begin by solving the  model for costs at which the agent would never search more than twice. This special case of the general problem illustrates how the agent hedges against uncertainty over the shape of the mapping by her choice of initial search location. It also illustrates for what quality realizations the agent continues or stops, highlighting the key tradeoff between the quality and informativeness of discoveries. Extending the ideas developed here, \Cref{sec:general} solves the general case where the agent may face lower costs and search many times. 

\subsection{Relevant Costs}

First, we identify a range of costs for which the agent searches at least once and most twice under any quality index $q \in Q$.

\begin{claim}
    If $c \in (\frac{1}{4}, \frac{1}{2})$, the agent searches at least once and at most twice in any optimal search policy, for every $q \in Q$.
\end{claim}

\begin{proof}
If the agent explores item $1/2$ and stops, then her payoff is at least $1/2 - c > 0$ for any $q \in Q$. Therefore, the agent searches at least once.

Next note that if the agent searches item $1/4$ and next searches item $3/4$, she can guarantee herself a payoff of at least $3/4 - 2c$. Therefore, a lower bound on the agent's payoff if she explores twice is $3/4 - 2c$.
An upper bound on the agent's payoff if she explores $k\geq 3$ times is $1 - kc$. Because $1/4 < c$, we have $1 - kc \leq 1 - 3c < 3/4 - 2c$. Therefore, the agent never searches more than twice in an optimal search policy. 
\end{proof}

Henceforth in this section we assume $c \in (\frac{1}{4}, \frac{1}{2})$.

\subsection{Bifurcation Risk}

\begin{figure}[!h]
    \centering
    \begin{subfigure}{.49\textwidth}
        \includegraphics{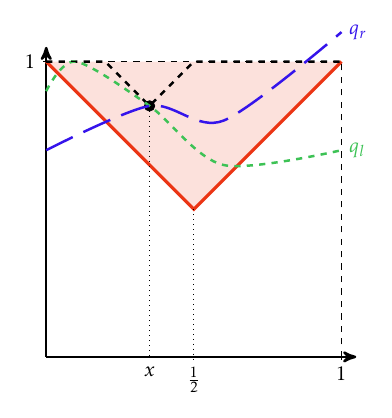}
    \end{subfigure}
    \begin{subfigure}{.49\textwidth}
        \includegraphics{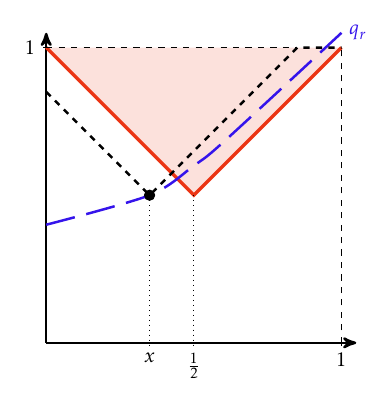}
    \end{subfigure}
    \caption{The agent in the first figure explores $x$ and faces bifurcation risk. Given what she knows, it is possible that target locations exist either exclusively to the left (e.g., if $q=q_l$) or right of her initial search (e.g., if $q = q_r$), and she guesses the wrong side to search next. The agent in the second figure does not face bifurcation risk. Due to the Lipschitz constraint, any feasible $q$ must lie on or below the dashed black line. Therefore, target locations must exist exclusively near the right end of the search space.}
    \label{fig:bifurcation}
\end{figure}

An agent faces \textit{bifurcation risk} if she is at a history where the search window consists of two disjoint intervals. This happens if her initial search, $x$, is of sufficiently high quality, i.e., $q(x) \geq \max\{x,1-x\}$, but is below the quality standard, i.e., $q(x) < 1$. A discovery of this quality leaves open the possibility that targets exist either to the left or right of the initial search; see the left panel of \Cref{fig:bifurcation}.

\begin{claim}\label{bifurcation}
    An optimal search policy concludes when the agent faces bifurcation risk.
\end{claim}

\begin{proof}
    Suppose the agent faces bifurcation risk after searching at $x$. Let 
    \[q_l(y) = \begin{cases}
        q(x) \text{, if } y \leq x\\
        \min\{q(x) + y - x, 1\} \text{, if } $y > x$.
    \end{cases}\]
    Note that $q_l$ is feasible at this history. If $q = q_l$, the agent's payoff if she searches to the left of $x$ and then concludes is $q(x) - 2 c < q(x) - c$. If she searches to the left of $x$, then at 1 and then concludes, she gets a payoff of $1 - 3c < 1/2 - c \leq q(x) - c$. Therefore, stopping at $x$ improves on the agent's worst case payoff to searching to the left of $x$, regardless of what she does afterward. Symmetrically, concluding search is also better than exploring to the right of $x$.
\end{proof}

When facing bifurcation risk, the agent finds herself at a crossroads when deciding where to search next. \Cref{bifurcation} shows that in the worst case, the direction she picks leads to a poor outcome, revealing that high quality discoveries instead exist on the other side of the search space.  Therefore, her optimal action at such a history is to conclude search.

To hedge against bifurcation risk, the searcher would explore close to the endpoints $0$ or $1$: bifurcation risk is then only possible if the searcher discovers a very high quality.

\subsection{Directional Risk}
\begin{figure}[!h]
    \centering
    \begin{subfigure}{.49\textwidth}
        \includegraphics{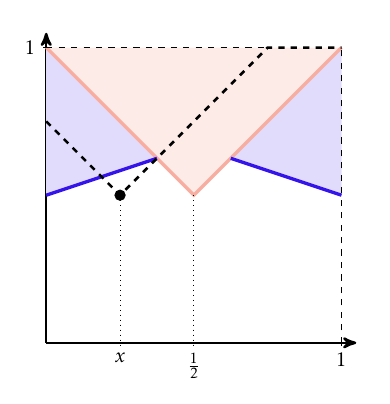}
    \end{subfigure}
    \begin{subfigure}{.49\textwidth}
        \includegraphics{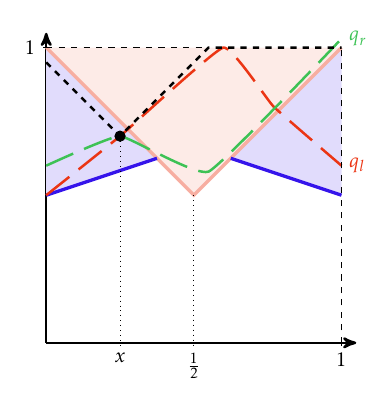}
    \end{subfigure}
    \caption{The agent explores $x$ and in the first figure she discovers a low quality $q(x)$. Lipschitz continuity then narrows down the possible target locations on the right end of the interval. Instead, in the second figure the agent discovers a high quality, which does not help significantly in narrowing down the location of the target.}
    \label{fig:directional}
\end{figure}

If the agent searches too close to $0$ or $1$, she risks having searched on the wrong side of the interval. In such a scenario, discovering a sufficiently low quality is valuable, because it narrows down the location of the target. 
This scenario is depicted on the left panel of \Cref{fig:directional}. Instead, if the
quality she discovers is sufficiently high, the search region remains large and an additional search is not profitable in the worst case. The right panel of \Cref{fig:directional} shows such a scenario. The agent faces \emph{directional risk} if she has searched once at $x$, the search window is contiguous and $ q(x) \geq \min\{\frac{2}{3}(1-c) + \frac{x}{3}, \frac{2}{3}(1-c)-\frac{x-1}{3}\}$. \Cref{fig:directional} highlights such region in blue.

\begin{claim}\label{claim:directional}
    An optimal search policy concludes when the agent faces directional risk.
\end{claim}
\begin{proof}
    Suppose the agent discovered a quality such that she is facing directional risk. 
    The worst-case quality index is similar to the one pictured in the right panel of \Cref{fig:directional}.
    If the first search was to the left of $\frac{1}{2}$, the agent is left with a search window of measure $q(x) - x$. Since the agent will optimally search at most twice, the worst-case quality she anticipates over this region is $1-\frac{q(x)-x}{2}$. She will search for a second time only if this quality minus the cost of search exceeds the quality $q(x)$ she already discovered. That is, 
    \begin{align*}
        q(x) < 1-\frac{q(x)-x}{2} - c \iff q(x) < \frac{2}{3}(1-c) + \frac{x}{3}.
    \end{align*}
    This is a contradiction, since the agent faces directional risk only if the quality she discovered is $q(x) \geq \min\Big\{\frac{2}{3}(1-c) + \frac{x}{3}, \frac{2}{3}(1-c)-\frac{x-1}{3}\Big\} =\frac{2}{3}(1-c) + \frac{x}{3}$.
    Similarly, if the first search is to the right of $\frac{1}{2}$, she is left with a search region of measure $x+q(x)-1$, and thus an expected worst-case quality of $\frac{3-x-q(x)}{2}$. Then, she will search a second time only if 
    \begin{align*}
        q(x) < \frac{3-x-q(x)}{2} - c \iff q(x) < \frac{2}{3}(1-c) - \frac{x-1}{3},
    \end{align*}
    which is again a contradiction.
\end{proof}

\subsection{Optimal Search Policy}

\begin{figure}[!h]
\centering
    \includegraphics{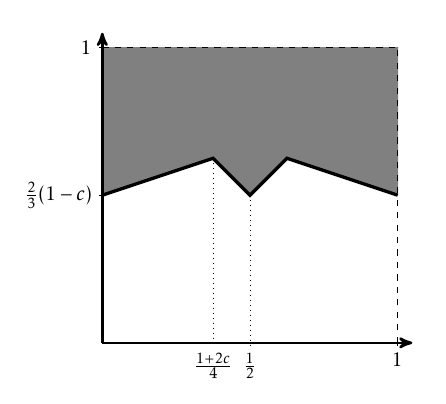}
    \caption{The stopping region, depicted in grey, is where the agent experiences either bifurcation or directional risk. The continuation region, in white, lies below it. The lower envelope of the stopping region, the solid black line, represents the worst-case discovery for any first period search.}
    \label{fig:bothrisks}
\end{figure}

The proofs of \Cref{bifurcation} and \Cref{claim:directional} show that it is optimal for the agent to stop in the second period if and only if she faces bifurcation or directional risk. \Cref{fig:bothrisks} identifies the stopping region and the continuation region.

Upon searching any item $x$ in the first period, and conditional on discovering a quaity that leaves the agent in the stopping region, her payoff is minimized when she discovers a quality at the bottom of this region. This is immediate.

Upon searching any item $x$ in the first period, and conditional on discovering a quality that leaves the agent in the continuation region, her payoff is minimized when she discovers a quality that puts her near the top this region. This is because the optimal policy continues to search in period two only when the second search is guaranteed to be of sufficiently high quality. This guarantee is smaller when the search window is larger, as there is more scope for the target to be further from the second search. The search window is largest when the searcher discovers an intermediate quality that barely puts the searcher in the continuation region.

Putting these observations together, the worst-case payoff for the agent, after any search $x$ in the first period, is discovering a quality $q(x)$ such that $(x, q(x))$ is on the lower envelope of the stopping region, i.e., the upper envelope of the continuation region. This is an $M$-shaped curve, as seen in \Cref{fig:bothrisks}. 

Therefore, a policy that maximizes the agent's worst-case payoff searches at one of the two peaks of the $M$-shaped curve in the first period. These peaks occur where the directional and bifurcation regions meet, so they are the solutions to the following system of equations:
\[
\begin{cases}
    q(x) = \min\Big\{ \frac{2}{3}(1-c) + \frac{x}{3}, \frac{2}{3}(1-c) - \frac{x-1}{3}\Big\}, \\
    q(x) = \max \{x, 1-x \},
\end{cases}
\]
which implies that in an optimal policy, $x_0 \in \Big\{\frac{1}{4} + \frac{c}{2}, \frac{3}{4} - \frac{c}{2} \Big\}$.

In the second period, the optimal policy stops if $(x_0, q(x_0))$ is in the stopping region. Otherwise it searches at the center of the remaining (contiguous) search window, as this would minimize the distance to the target item in the worst case. More formally:

\begin{claim}\label{claim:optimal}
    Let $x_0 \equiv \frac{1}{4} + \frac{c}{2}$. Let $$\sigma(h_0) = x_0,$$ and for any $h_1 \equiv (x_0, z_0)$, let 
    \[\sigma(h_1) = \begin{cases}
        \emptyset &\text{ if } z_0 \geq \frac{3}{4} - \frac{c}{2} \\
        1 - \frac{l_{h_1}}{2} &\text{ otherwise.}
    \end{cases}\]
    Then $\sigma$ is an optimal policy.\footnote{While \Cref{claim:optimal} characterizes a left-to-right optimal policy, there is a symmetric right-to-left policy as well, so the optimal policy is not unique.}
\end{claim}
 
Intuitively, the worst-case outcome for the agent is one where she discovers an intermediate quality in the first period. This quality is barely high enough to stop search in the first period and not low enough to better hone in on the target in the second period. To maximize payoffs in such a scenario, the agent searches at a location where she is equally well hedged against bifurcation risk and directional risk.

\subsection{Simultaneous vs Sequential Search} 

The optimal policy accounts for the option value of searching a second time if the first discovery is low quality.
To see how, it is instructive to compare an optimal policy in our sequential search model to the optimal policy in a simultaneous search model.

If the agent had to pick two locations at once to search simultaneously, she would choose locations $1/4$ and $3/4$. This configuration ensures an item of quality 1 cannot be farther than distance $1/4$ from one of the chosen locations, guaranteeing that the agent gets a payoff of at least $3/4 - 2c$.

With sequential search, there is option value to deciding whether or not to continue with a costly second search. The agent would no longer search again upon finding a quality of $3/4$. This in turn affects the worst-case outcome of searching at $x=1/4$ in the first place. 
The first search of the optimal sequential policy cleaves closer to the center. Getting a lower draw is more informative when searching closer (but not exactly at) the center, because it generates more information for a second search. Under the optimal policy, the agent gets a payoff of at least $3/4 - 1.5c$ (see \Cref{fig:bothrisks}).

\section{Proof of the General Result}\label{sec:general}

We convey the key ideas for the proof of \Cref{thm:opt} in the general case. 
For any strategy $\sigma\in \Sigma$, the continuation value at any history $h\in H$ is \[V^\sigma_q(h) \equiv p(h^{\sigma}_q) - (p(h) - z^*_h).\]
First, we argue that at a non-terminal ordered search history $h$ and conditional on following $\sigmalr$, the quality that minimizes the agent's continuation payoff after searching at $\sigmalr(h)$ is $\phi(l_h)$:

\begin{lemma}\label{lemma:threshold_is_worst}
    For any non-terminal $h\in H^{\sigmalr}$ and for all $q \in Q_h$,  
    $$\phi(l_h) - c \leq V^{\sigmalr}_q(h).$$
\end{lemma}

Clearly, the agent is better off if she discovers a quality above the threshold, as $\phi(l_h)$ is the lowest quality for which the policy recommends stopping. 
The argument that discovering any quality below the threshold improves the agent's eventual payoff upon stopping is less immediate. On the one hand, discovering low quality is \emph{informative}: the search window shrinks, narrowing the location of target items. On the other hand, exploiting this information requires additional costly searches.

\begin{figure}
    \centering
    \includegraphics[width=0.9\linewidth]{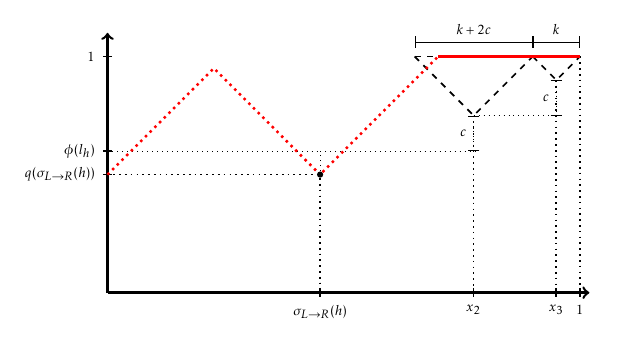}
    \caption{When the agent discovers a quality $q(\sigmalr(h)) < \phi(l_h)$, the search window shrinks to the solid red region in this image, with size $l_{h^{+1}_q}$ less than $k + (k+2c)$. Suppose the agent chose to search locations $x_2$ and $x_3$ sequentially. If she discovered worse qualities than those at the vertices of the respective triangles, the search window would be empty, which is a contradiction. By searching at the peaks of the triangles, the agent must then be able to secure a continuation utility $V_q^\sigma(h)$ larger than $\phi(l_h)- c$.}
    \label{figure:triangles_cover_smaller_S}
\end{figure}

To understand why the tradeoff always goes in the agent's favor, recall the geometric depiction of $\sigmalr$ at an ordered search history $h$ in \Cref{figure:triangles}. 
Suppose searching at $\sigmalr(h)$ reveals a quality lower than $\phi(l_h)$. The search window at this new history, $h^{+1}_q$, shrinks to a region smaller than the interval covered by the triangles centered at $x_2$ and $x_3$, shown in \Cref{figure:triangles_cover_smaller_S}. At $h^{+1}_q$, consider the `non-responsive' strategy $\sigma$ that searches at $x_2$, stops if $x_2\geq \phi(l_h) + c$, and searches and stops at $x_3$ otherwise. This strategy nets the agent either $\phi(l_h) + c - c = \phi(l_h)$ after the first search or $\phi(l_h) + 2c - 2c = \phi(l_h)$ in the second search. A discovery of quality lower than $\phi(l_h) + 2c$ is impossible if $x_3$ is searched on path: the Lipschitz bounds would imply that the search window is empty at such a history, constradicting the fact that a target item exists. Strategy $\sigmalr$ performs even better than the non-responsive strategy at history $h^{+1}_q$ by adapting the search location to the smaller search window, proving \Cref{lemma:threshold_is_worst}.

\begin{figure}
    \centering
    \includegraphics[width=0.9\linewidth]{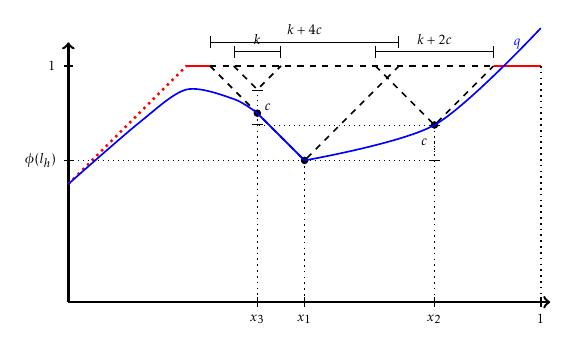}
    \caption{Searching in $x_1 \neq \sigmalr(h)$, $x_2$, and $x_3$ leaves a non-empty search window. Then, there exist quality indices that pass through the items discovered and attain the target. The agent will never do better than the threshold $\phi(l_h)$.}
    \label{figure:triangles_not_partition}
\end{figure}

Next we argue that at any ordered search history, the agent can do no better by using a strategy other than $\sigmalr$:
\begin{lemma}\label{lemma:no better with other sigma}
   For any non-terminal $h\in H^{\sigmalr}$ and $\sigma \in \Sigma$, there is a $\tilde{q} \in Q_h$ with $$V_{\tilde{q}}^\sigma(h)\leq \phi(l_h) - c.$$ 
\end{lemma}

Suppose $\sigma$ is a strategy that always searches inside the search window in any period.
We construct a feasible quality index $\tilde{q}$ such that the latest discovery at any step is $c$ better than the previous.\footnote{If $\sigma$ searches outside the search window, the greatest feasible improvement in quality is even smaller; compare $q(x_3)$ and $q(x_2)$ in \Cref{figure:triangles_not_partition}.} 
Such a quality index at most maintains the agent's continuation value constant, proving that $\phi(l_h)-c$ is an upperbound to the agent's belief about her continuation payoff. Interestingly, the threshold is such that the agent will never search more than $N(c,l_h)$ items, thus making $N(c,1)$ the effective game horizon (and guaranteeing that any optimal strategy must terminate).

\section{Conclusion}

Running a mile under 4 minutes was a feat that was seriously attempted by athletes since the 1880s. For many decades, it was unclear whether the barrier to achieving this feat were physical or psychological. However, in the 1954, Roger Bannister, an unlikely and iconoclastic runner who eschewed the training wisdom of the day finally ran a mile in under 4 minutes. Several other soon followed him, though he did not immediately reveal his training regimen to his competitors. Today, it is an impressive but not rare feat. A tribute to Bannister in the \textit{Harvard Business Review} writes (\citet{hbr2018}):

\begin{quote}
``\ldots what goes for runners goes for leaders running organizations. In business, progress does not move in straight lines. Whether it’s an executive, an entrepreneur, or a technologist, some innovator changes the game, and that which was thought to be unreachable becomes a benchmark, something for others to shoot for. That’s Roger Bannister’s true legacy\ldots"
\end{quote}

Anecdotes like these suggest that simply knowing that some target is achievable affects how agents look for it.
We model rediscovery and find that having a target turns search into a process of elimination. Agents are emboldened by failure: rather than stop after a disappointing discovery, they set their sights higher.

Insofar as rediscovery is a source of innovation, it may be of interest to policymakers to understand it, and our model may serve as a helpful starting point.
Policymakers may, for example, be interested in industry-level innovation dynamics of firms that carry on their research in private, e.g. by keeping their innovations as trade secrets. This may be captured by studying a sequence of rediscovery attempts by firms who observe only the final outcomes but not failed attempts of their predecessors. 

Policymakers may also be interested in encouraging rediscovery. One may use our framework to study whether this is better done by subsidizing search, rewarding good discoveries or a mixture of the two. Similarly, platforms that mediate search and learning may be interested in designing recommendations to influence consumer search. Our model may serve as a helpful baseline for search in the absence of such influence. We leave these as topics for further study.

\newpage

\bibliographystyle{ACM-Reference-Format} 
\bibliography{references}

\appendix
\section{Proofs}\label{appendix:proofs}

Recall that all proofs are developed in continuation value space. Unless necessary to avoid confusion, we drop the dependence of reachable histories $h^{+k}_q(\sigma)$ on the policy $\sigma$.

\subsection*{Proof of \Cref{lemma:terminal or ordered}.}

Because $h$ is non-terminal, $\sigmalr(h) = 1 - l_h + 1 - \phi(l_h)$. We consider two cases. 

\medskip
\noindent

\textit{Case 1:} $q(\sigmalr(h)) \geq \phi(l_{h})$.
Then, 
$$S_{h^{+1}_q} = [1-l_h, \sigmalr(h) - 1 + q(\sigmalr(h))] \cup [\sigmalr(h) + 1 -q(\sigmalr(h)), 1],$$
where, by the assumption in this case, \begin{align*}
    \sigmalr(h) - 1 + q(\sigmalr(h)) &\geq \sigmalr(h) - 1 + \phi(l_h)\\ &= 1-l_h + 1 - \phi(l_h) - 1  + \phi(l_h)\\ &= 1-l_h.
\end{align*}
Therefore, $h^{+1}_q$ is not an ordered search history. Let $n = N(c,l_h)$. Then, the length of the search window is
\begin{align*}\label{equation:relationship lengths}
\begin{split}
   l_{h^{+1}_q} &> q(\sigmalr(h)) - \sigmalr(h) \\ &= 2(\phi(l_h)) - 2 + l_h \\
   &= \frac{n-1}{n}l_{h} - (n-1)c.
   \end{split}
\end{align*}
Because $\phi(\cdot)$ is decreasing, $\phi(l_{h^{+1}_q}) <  \phi\big(\frac{n-1}{n}l_{h} - (n-1)c\big)$. By Equation \ref{eq:N(c,a)}, $n(n-1)c \leq l_h < n(n+1)c$. This implies that $(n-1)(n-2)c \leq \frac{n-1}{n}l_{h} - (n-1)c < (n-1)nc$. Therefore, 
\begin{align*}
    \phi\Big(\frac{n-1}{n}l_{h} - (n-1)c\Big) &= 1 - \frac{1}{2(n-1)}\Big(\frac{n-1}{n}l_{h} - (n-1)c\Big) - \frac{n-2}{2}c \\
    &= 1 - \frac{l_h}{2n} - \frac{n-1}{2}c + c\\
    &= \phi(l_h) + c.
\end{align*}
Therefore, $$\phi(l_{h^{+1}_q}) -c < \phi(l_h) \leq q(\sigmalr(h)).$$ Because $\sigmalr$ ignores past discoveries, this directly implies that $h^{+1}_q$ is terminal according to \Cref{definition:opt_policy}.

\medskip
\noindent

\textit{Case 2:}
Suppose instead that $q(\sigmalr(h)) < \phi(l_{h})$. Then, $$S_{h^{+1}_q} = [\sigmalr(h) + 1 - q(\sigmalr(h)), 1],$$ so $h^{+1}_q$ is an ordered search history. 
By assumption 
$l_{h^{+1}_q} < \frac{n-1}{n}l_{h} - (n-1)c$, so $\phi(l^{+1}_h) > \phi(l_h) + c$. 
Then, combining, we get $q(\sigmalr(h)) < \phi(l_{h}) < \phi(l^{+1}_h) - c$, therefore $h^{+1}_q$ is not a terminal history.  \hfill \openbox

\subsection*{Proof of \Cref{prop:sigmalr is simple}}
    
    First, the left-to-right policy is \emph{threshold} by definition. Next, the threshold depends on the history only through the length of the search window $l_h$, so $\sigmalr$ is \emph{index}.

    Third, note that by equation \ref{eq:N(c,a)},
    \begin{align*}
    \phi(l) = 1-\frac{l}{2N(c,l)} - \frac{N(c,l)-1}{2}c &\leq 1-\frac{N(c,l) (N(c,l)-1)}{2} - \frac{N(c,l)-1}{2}c \\
    &= 1-(N(c,l)-1)c \leq 1
    \end{align*}
    as long as $N(c,l) > 0$. If $h^{+1}_q$ is not a terminal node, then $l_{h^{+1}_q} \leq l_h - 2(1-\phi(l_h))$, so 
    \begin{align*}
        \sigmalr(h^{+1}_q) > 1 - l_{h^{+1}_q} &\geq 1 - l_{h} +  (2 - \phi(l_h))\\ &> 1 - l_{h} + 1 - \phi(l_h) = \sigmalr(h),
    \end{align*}
    where the last inequality follows from the fact that $\phi(l) \leq 1$ for any $l$. Therefore, the left-to-right policy is \emph{directional}.

    Finally, \Cref{lemma:terminal or ordered} implies that if $h^{+1}_q$ is not a terminal history, it is an ordered search history. The search window shrinks from $h$ to $h^{+1}_q$ for any $q\in \Omega_h$ where the discovery has quality below the target, i.e., $l_{h^{+1}_q} < l_h$. Because $\phi(\cdot)$ is decreasing, $\phi(l_{h^{+1}_q}) > \phi(l_h)$, so the stopping threshold increases between $h$ to $h^{+1}_q$. Increasing stopping thresholds imply that the left-to-right policy \emph{ignores past discoveries}.

\subsection*{Proof of \Cref{thm:opt}}
To show that $\sigmalr$ is optimal, we must consider three kinds of deviations:
\begin{enumerate}
    \item Stopping when $\sigmalr$ recommends continuing, 
    \item Continuing when $\sigmalr$ recommends stopping, 
    \item Searching an item other than $\sigmalr(h)$.
\end{enumerate}
\Cref{lemma:threshold_is_worst} shows that deviations of the first or second kind do not strictly improve the agent's worst-case eventual payoff. \Cref{lemma:no better with other sigma} shows that deviations of the third kind also do not strictly improve the agent's worst-case eventual payoff. \openbox

\subsection*{Proof of \Cref{lemma:threshold_is_worst}}

Let $h\in H^{\sigmalr}$ be a non-terminal history and $q\in Q_h$.

\medskip
\noindent
\textit{Base Case:} $N(c,l_h) = 1$. Because $h$ is non-terminal, $S_h = [a,1]$ for some $a\in(0, 1]$, so $\sigmalr(h) = a + \frac{l_h}{2}$. The threshold $\phi(l_h)$ is $1-\frac{l_h}{2}$. There is no feasible $\tilde{q} \in Q_h$ such that $\tilde{q}(\sigmalr(h)) < \phi(l_h)$, because such a $\tilde{q}$ does not attain a value of $1$ anywhere. Therefore, $q(\sigmalr(h)) \geq \phi(l_h)$, so $\sigmalr$ stops, so $\phi(l_h)-c\leq V_q^{\sigmalr}(h)$.

\medskip
\noindent
\textit{Induction Hypothesis:} Suppose that for any $h'\in H^{\sigmalr}$ such that $N(c,l_{h'}) \leq n-1$ and for all $q\in\Omega_{h'}$, $$\phi(l_{h'}) - c \leq V_{q}^{\sigmalr}(h').$$

\medskip
\noindent
\textit{Inductive Step:} $N(c,l_h) = n$. The case where $q(\sigmalr(h)) \leq \phi(l_h)$ is trivial, so suppose that $q(\sigmalr(h)) < \phi(l_h)$.
Then, $$S_{h^{+1}_{q}} = [\sigmalr(h) + 1 - q(\sigmalr(h)), 1].$$ Then
\begin{align*}
    l_{h^{+1}_{q}} = q(\sigmalr(h)) - \sigmalr(h)  &< \phi(l_h) - \sigmalr(h) \\ 
    &= \phi(l_h) - (1 - l_h + 1 - \phi(l_h)) \\
    &= 2\phi(l_h) + l_h - 2 \\
    &= \frac{n-1}{n}(l_h - nc) \\
    & < n(n-1)c,
\end{align*}
where the last inequality is because $l_h < n(n+1)c$ by Equation \ref{eq:N(c,a)}. This implies that $N(c,l_{h^{+1}}(q)) = n-k$, for some $1\leq k < n$. By the inductive hypothesis, $$V_q^{\sigmalr}(h^{+1}_{q}) \geq \phi(l_{h^{+1}_{q}}) -c.$$ Moreover, $l_{h^{+1}_{q}} < (n-k)(n-k + 1)c$ by Equation \ref{eq:N(c,a)}. Because $\phi$ is decreasing, \begin{align*}
    \phi(l_{h^{+1}_{q}}) - c \geq \phi\big((n-k)(n-k+1)c\big) - c &= 1-\frac{(n-k)(n-k+1)c}{2(n-k)} - \frac{(n-k-1)}{2}c -c\\
    &= 1 - (n-k)c - c
\end{align*}
Finally, because $l_h \geq n(n-1)c$ by Equation \ref{eq:N(c,a)}, 
\[\phi(l_h) \leq 1-\frac{n(n-1)}{2n}c - \frac{n-1}{2}c = 1 - (n-1)c \leq 1 - (n-k)c - c.\]
Therefore, $V_q^{\sigmalr}(h^{+1}_{q}) \geq \phi(l_h)$. Because $h$ is non-terminal, $V_q^{\sigmalr}(h) = V_q^{\sigmalr}(h^{+1}_{q}) - c$, so $$V_q^{\sigmalr}(h) \geq \phi(l_h) -c.$$ \hfill \openbox

\subsection*{Proof of \Cref{lemma:no better with other sigma}}

For $k=0, 1, 2, 3\ldots$ define history $h^{+k}$ as follows:

$$h^{+k} \equiv
\begin{cases*}
    h, \text{ if } k=0\\
    h^{+(k-1)}, \text{ if } \sigma(h^{+(k-1)}) = \emptyset\\
    h^{+(k-1)} \cup \big(\sigma(h^{+(k-1)}\big), \ z^*(h^{+k})\big), \text{ otherwise,}
\end{cases*}$$
where,
$$z^*(h^{+k}) = \min \Big\{ \phi(l_h) + (k-1)c, \max_{q \in Q_{h^{+(k-1)}}} q\big(\sigma(h^{+(k-1)})\big) \Big\}$$.

For all $k\geq 1$, the best item at $h^{+k}$ has quality at most $\phi(l_h) + (k-1)c$. The agent had to pay $kc$ to reach history $h^{+k}$ from $h$. 
Therefore, letting $\bar{k}$ be such that $h^{+\bar{k}}$ is a terminal history, for any $\tilde{q} \in Q_{h^{+\bar{k}}}$,
\[V^\sigma_{\tilde{q}}(h) \leq \phi(l_h) + (k-1)c - kc = \phi(l_h) - c.\]

It remains to be shown that there exists some $\tilde{q}\in Q_{h^{+\bar{k}}}$. For this, it suffices to show that (i) for any two $(x', z'), (x'', z'') \in h^{+\bar{k}}$, $\frac{z''-z'}{x''-x'}\leq L$, and (ii) $S_{h^{+\bar{k}}}$ is nonempty. The former follows directly by construction of $h^{+\bar{k}}$, so next we argue the latter.

Let $n = N(c,l_h)$. Note that $S_{h^{+k}} =  S_{h^{+(k-1)}}$ if $z^*(h^{+k}) < \phi(l_h) + (k-1)c$. 
Therefore, for $k>n$, either  
\begin{align*}z^*(h^{+k}) = \phi(l_h) + kc > \phi(l_h) + nc &= 1 - \frac{l_h}{2n} - \frac{n-1}{2}c + nc \\&> 1 - \frac{n(n+1)}{2n} - \frac{n-1}{2}c + nc = 1,\end{align*} which would imply that $S_k$ is nonempty,
or $S_{h^{+k}} =  S_{h^{+(k-1)}}$. 

Then, by induction, $h^{+\bar{k}}$ is nonempty if $\bar{k}>n$ and $S_{h^{+n}}$ is nonempty. Otherwise, if $\bar{k} < n$, $S_{h^{+n}} \subseteq S_{h^{+\bar{k}}}$. Therefore, to show that $h^{+\bar{k}}$ is nonempty, it suffices to show that $S_{h^{+n}}$ is nonempty.

From the $h^{+(k-1)}$ to $h^k$, the search window shrinks by at most an open ball of radius $1-z^*(h^{+k})$, denoted by $B(h^{+k})$. 
The measure of the search window $S_{h^{+n}}$ must then be
\begin{align*}
    \mu(S_{h^{+n}}) \geq \mu\bigg(S_h \setminus \cup_{k=0}^{n-1} B(h^{+k})\bigg) &\geq \mu(S_h) - \sum_{k=0}^{n-1} 2\big(1-z^*(h^{+k})\big) \\
    &\geq l_h - \sum_{k=0}^{n-1} 2\big(1-\phi(l_h) - kc\big) \\
    &= l_h - \sum_{k=0}^{n-1} 2\Big(
    \frac{l_h}{n} + \frac{n-1}{2}c - kc\Big) \\
    &= l_h - l_h + \frac{n(n-1)}{2}c - \frac{n(n-1)}{2}c = 0.
\end{align*}
 
The search window at history $h^{+n}$ has always weakly positive measure. If the  measure is strictly positive, $S_{h^{+n}}$ is nonempty. Otherwise, note that since each ball $B(h^{+k})$ is an open set, the search window must contain at least an isolated point.\hfill \openbox

\end{document}